\documentclass[manuscript]{aastex}
\usepackage{listings}
\usepackage{url}

\newcommand{\autofit}{{\it autofit\ }}

\shorttitle{DR7 White Dwarf Catalog}
\shortauthors{Kleinman et al.}

\begin{document}
\title{SDSS DR7 White Dwarf Catalog}

\author{S.J. Kleinman\altaffilmark{1}}
\and
\author{S.O. Kepler\altaffilmark{2}, 
D. Koester\altaffilmark{3},
Ingrid Pelisoli\altaffilmark{2},
Viviane Pe\c{c}anha\altaffilmark{2},
A. Nitta,\altaffilmark{1},
J.E.S. Costa\altaffilmark{2},
J. Krzesinski\altaffilmark{4},
P. Dufour\altaffilmark{5}, 
F.-R. Lachapelle\altaffilmark{5}, 
P. Bergeron\altaffilmark{5}, 
Ching-Wa Yip\altaffilmark{6},
Hugh C. Harris\altaffilmark{7},
Daniel J. Eisenstein\altaffilmark{8},
L. Althaus\altaffilmark{9}, 
A. C\'orsico\altaffilmark{9}
}

\altaffiltext{1}{Gemini	Observatory, 670 N. A'ohoku Place, Hilo HI 96720,
USA}
\altaffiltext{2}{Instituto de F\'{\i}sica, Universidade Federal do Rio Grande do Sul, Porto Alegre, RS, Brazil}
\altaffiltext{3}{Institut f\"ur Theoretische Physik und Astrophysik, Universit\"at Kiel, 24098 Kiel, Germany}
\altaffiltext{4}{Mt. Suhora Observatory, Pedagogical University of Cracow, ul. Podchorazych 2, 30-084 Cracow, Poland}
\altaffiltext{5}{D\'epartement de Physique, Universit\'e de Montr\'eal, C. P. 6128, Succ. Centre-Ville, Montr\'eal, Qu\'ebec H3C 3J7, Canada}
\altaffiltext{6}{Department of Physics and Astronomy, The Johns Hopkins
University, 3701 San Martin Drive, Baltimore, MD 21218, USA}
\altaffiltext{7}{United States Naval Observatory, Flagstaff Station, 10391 West Naval Observatory Road, Flagstaff, AZ 86001-8521, USA; hch@nofs.navy.mil}
\altaffiltext{8}{Harvard Smithsonian Center for Astrophysics, 60 Garden
St., MS \#20, Cambridge, MA 02138, USA}
\altaffiltext{9}{Facultad de Ciencias Astron\'omicas y Geof\'{\i}sicas, Paseo del Bosque S/N, (1900) La Plata,  Argentina}

\setcounter{footnote}{0}

\begin{abstract}
We present a new catalog of spectroscopically-confirmed white dwarf
stars from the Sloan Digital Sky Survey Data Release 7 spectroscopic
catalog. 
We find 20\,407 white dwarf spectra, representing 19\,712 stars,
and provide atmospheric model fits to
14\,120 DA and 1011 DB white dwarf spectra from 12\,843 and 923 stars, respectively.
These numbers represent a more than factor of two increase
in the total number of white dwarf stars from the previous SDSS white dwarf 
catalog based on DR4 data.
Our distribution of subtypes varies from previous catalogs
due to our more conservative, manual classifications of each
star in our catalog,  supplementing our automatic fits.  In particular, we find
a large number of magnetic white dwarf stars whose small Zeeman splittings
mimic increased Stark broadening that would otherwise result in an
overestimated
$\log{g}$ if fit as a non-magnetic white dwarf.
We calculate mean DA and DB masses for our clean, non-magnetic sample and find
the DB mean mass is statistically larger than that for the DAs.
\end{abstract}


\section{Introduction}
The Sloan Digital Sky Survey \citep[SDSS,][]{yor00} has had impacts in
astronomy far beyond its main mission of exploring the three-dimensional
structure of our Universe.  The Sloan Extension for Galactic Exploration and 
Understanding survey \citep[SEGUE,][]{lee08}, part of the
second-generation SDSS surveys, extended the survey's mission to
unraveling the nature and structure of our own Milky Way galaxy by
targeting mostly new fields in and around the Galactic disc.  As a result,
the spatial distribution of the new (largely Galactic disc) white dwarf stars 
in this catalog will be very different from those in the earlier
catalogs from SDSS Data Releases 1 \citep{kle04} and  4 \citep{eis06} where the
focus was on extragalactic objects and the Galactic disc was purposefully 
avoided.
The study of white dwarf stars has benefited greatly from the increased number 
of objects provided by the SDSS, there being 66 papers between 2005 and 2012 
reported by the SAO/NASA Astrophysics Data System, for example,  with {\it SDSS} 
and {\it white dwarf} in the title.   Numerous other papers refer to
SDSS-discovered white dwarf stars without indicating so in the title.
The first full white dwarf catalog from SDSS data
\citep{kle04}, based on SDSS Data Release 1 \citep[DR1,][]{dr1},
roughly doubled the number of then known white dwarf stars.
Using data from the SDSS Data Release 4 \citep[DR4,][]{dr4}, \citet{eis06} 
reported over 9\,000 spectroscopically-confirmed white dwarfs stars from the SDSS,
again roughly doubling the combined number of white dwarf stars known after
SDSS DR1.    With the release of Data Release 7 from the SDSS
\citep[DR7,][]{dr7}, we again roughly double the
number of identified white dwarf stars compared to those in the DR4 sample.  

The first release of SEGUE data started in SDSS Data Release 6 \citep{dr6},
with more released in SDSS Data Release 7 \citep{dr7}.  In the original
SDSS Survey, white dwarf spectra were obtained primarily as a bi-product of other 
high-priority categories of targets.  Almost all were hot white dwarf stars because white dwarf
stars cooler than $\approx$ 7\,000K have colors similar to the more numerous FGK main sequence stars which 
were specifically not targeted.
Most of
the white dwarf stars in the survey were not targeted for
spectroscopy as white dwarf star candidates and were instead rejects 
from targeting algorithms for other kinds of objects; \citet{kle04, har03} discuss the
details of the DR1 target selection and the makeup of the white dwarf spectroscopic sample.
The SEGUE survey, however, specifically targets stars (see
\url{http://www.sdss3.org/dr8/algorithms/segueii/segue_target_selection.php})  and
cool white dwarf stars were effectively targeted for the first time using
their reduced proper motions.
The net result is that the number of white dwarf stars
observed per SDSS spectroscopic plate has remained roughly constant at
~25/plate through each SDSS Data Release, although the selection mechanism
is significantly different.

Here, we report on the white dwarf catalog built from the SDSS Data Release 7.
We have applied automated techniques supplemented by complete, 
consistent human identifications of each candidate white dwarf spectrum.  We make use of the latest SDSS reductions
and white dwarf model atmosphere improvements in our spectral fits,
providing $\log{g}$ and $T_\mathrm{eff}$ determinations for each identified clean DA
and DB where we use the word {\it clean} to identify spectra that 
show only features of non-magnetic, non-mixed, DA or DB stars.
Our catalog includes all white dwarf stars from the earlier
\citet{kle04} and \citet{eis06} catalogs, although occasionally with
different identifications, as discussed below.

Looking for infrared excesses around DA white dwarf stars, \citet{gir11}
use a photometric method to identify DA white dwarf stars with $g<19$
from the SDSS and find 4636 spectroscopically-confirmed DAs in DR7 with
another 5819 expected DAs in the photometric sample.  Our sample is not
magnitude-selected
(although classifications typically get more uncertain
for $g \approx 19.5$ and below) and includes DB and all other white dwarf
subtypes as well.   We do not, however, consider candidate white dwarf stars
that do not have SDSS spectra.

We note that although we did not fit white dwarf plus main sequence models
to our apparently composite spectra,
others \citep{sil06, mar11,gir11,koe11,deb11,ste11,reb12} have specifically studied these
spectra.

\section{Candidate Selection}
SDSS DR7 contains over 1.6 million spectra and we did not have the
facilities to fit and identify each spectrum.  We therefore had to extract a
smaller subsample of candidates from these spectra that we
could later fit and examine as possible white dwarf stars.
To form our candidate sample, we employed two different
techniques.  First, we reproduced the candidate selection from
\citet{eis06}, but implemented it completely within the SDSS DR7 Catalog Archive
Server (CAS)\footnote{http://cas.sdss.org/dr7} as an SQL query.  This query 
returned 24\,189 objects.

Second, we used the SDSS and SEGUE target classification and spectrum
analysis fields and selected any object that was either targeted as a
possible white dwarf star or whose spectrum was determined to likely be any 
kind of white dwarf star.  This query returned 48\,198 spectrum IDs.
Both queries are listed in full in
the Appendix and queried all available DR7 spectra via the specObjAll
table.

Combined, these queries resulted in 53\,408 unique spectra, of which 5209 uniquely
satisfied the \citet{eis06} criteria and 29\,218 uniquely satisfied the new
target/classification criteria.  18\,981 spectra satisfied both. Later, we
discovered that 4362 sky spectra made our sample (they should have been
explicitly excluded from our queries, but were not), so these were deleted
and the resulting sample size became 49\,046 spectra.  Ultimately, 
4\% of the objects which were selected only by the targeting criteria,
36\% of those selected only by the
\citet{eis06} criteria, and 90\% of those that satisfied both criteria
were labeled as white dwarf stars.  The combined set of criteria is a very
powerful way to identify white dwarf spectra in the SDSS, accounting for
83\% of our identified white dwarf stars.  Only 17\% of the identified white
dwarf stars, therefore, satisfied just one of the two selection criteria.

We further pared our sample by using lists of previously-identified
SDSS spectra.  We removed the known quasars \citep{sch10}, BL~Lacs \citep{plo10}, and
once run through our DA and DB model-fitting program, we fit the rejects
with galaxy and quasar templates \citep{yip04, yip04a} to remove these objects from further
consideration.
This process
removed an additional 6892 spectra, resulting in a final sample of
42\,154 spectra.  

\section{White Dwarf Atmosphere Models}

The mechanics of our \autofit fitting program, 
which fits the observed spectra to our synthetic model spectral grid by 
$\chi^2$ minimization,
remain the same as described in
\citet{kle04} and \citet{eis06}, but substantial improvements have been made to
our atmospheric model grid, both in the model physics and the
parameters of the grid itself.

We use updated Koester 
\citep{koe09,koe10} model atmospheres, 
with the following 
significant changes since \citet{kle04}:
\begin{itemize}
\item For the ten lowest Balmer and Lyman lines, the standard VCS
  tables \citep{lem97} were replaced with new tables calculated by
  \citet{tre09}. These calculations consistently include the
  Hummer-Mihalas occupation probability formalism into
  the profile calculation. 
\item The Stark broadening profiles from
  \citet{bea97} for hydrogen have been convolved with the
  neutral broadening profiles to add another dimension for the neutral
  particle density to the broadening tables. For the three lowest
  Balmer lines, we used the self-broadening data of \citet{bar00}.
  For the higher series members, we used the sum of resonance 
  \citep{ali65} and van der Waals broadening \citep{uns68}. For
  the helium lines, we used self-broadening data from 
\citet{leo95} and \citet{mul91}.
For the remaining lines,
  simple estimates for resonance and van der Waals broadening were
  used. 
\item The Holtsmark microfield distribution was replaced by the Hooper
  (1966, 1968) distribution using the approximations in Nayfonov et
  al.\ (1999). This distribution includes correlations between the
  charged perturber particles. The changes in the occupation
  probabilities for higher Balmer lines, where the occupation
  probability varies between 0.1--1.0, are quite significant.
\item For the DBs, we now use the $ML2/\alpha=1.25$ approximation.
  We find that this value best describes the
  location of the DBV instability strip 
  \citep{mon07,cor09,mon10}.  \citet{mon10}, in particular,
exclude values of $\alpha<0.8$ through their analysis of the
the convection zone for the pulsating DB white dwarf, GD~358 while
\citet{ber11} justify this value in their atmospheric modeling.
\item  For the DAs, we use $ML2/\alpha=0.6$. Note that with the use of
these improved Stark profiles, \citet{tre10}
showed that a slightly more efficient convective
energy transport with $\alpha=0.8$ should be used, although we feel our
atmospheric parameters with $\alpha=0.6$ are appropriate in the present
context.
\end{itemize}

Our model grid now extends to $\log{g}=10.0$ and is denser than that used
in \citet{eis06} and \citet{kle04}.  For DAs, the grid extends in $\log{g}$ 
from 5.0 to 10.0 in steps
of 0.25 while $T_\mathrm{eff}$ goes from 6\,000K to 10\,000K in steps of 250K, 10\,000K
to 14\,000K in steps of 100K, 14\,000K to 20\,000K in steps of 250K, 20\,000K to
50\,000K in steps of 1\,000K, and 50\,000K to 100\,000K in steps of 2\,500K.  The DB
grid runs from $\log{g}=$ 7.0 to 10.0 in steps of 0.25, with $T_\mathrm{eff}$
extending from 10\,000K to 18\,000K in steps of 250K and from 18\,000K to 50\,000K in
steps of 1\,000K.

\section{Spectral Fitting}

Once we completed our candidate list, we fit all 42\,154 candidate white
dwarf spectra and colors with our \autofit
code described in \citet{kle04} and \citet{eis06}.  
{\it Autofit} fits only clean DA and DB models, so does not
recognize other types of white dwarf stars.  In addition to the best
fitting model parameters, it also outputs a goodness of fit estimate and several
quality control checks and flags for other features noted in the spectrum
or fit.

We took the output from \autofit and separated the results into good
DA and DB fits (14\,271 spectra) and all else (27\,883 spectra).  We looked at all 
the good DA and DB fits to
verify they were indeed normal DAs and DBs and made about 1\,000 ID changes.
In almost all cases, we agreed each spectrum was one of
a DA or DB white dwarf star, but found they also contained additional spectral 
features
not fit by our models, resulting in new identifications like DAB,
DAH, DA+M, etc.
We also looked at each spectrum \autofit failed to classify as a DA or DB
dwarf star, identifying some as other white dwarf subtypes while most were simply other
non-white-dwarf stellar spectra.

\subsection{Spectral Classification}

Since \autofit can only classify clean DA and DB spectra, we knew we would
have to look at its rejected spectra for other white dwarf spectral
types.  
Because we were interested in obtaining accurate mass distributions for
our DA and DB stars, we were conservative in labeling a spectrum as a clean
DA or DB.  That is, we were liberal in adding additional subtypes
and uncertainty notations if we saw signs of other elements, companions, or
magnetic fields in the spectra.  While some of our mixed white dwarf
subtypes would probably be identified as clean DAs or DBs with better
signal-to-noise spectra,  few of our identified clean DAs or DBs would
likely be found to have additional spectral features within our detection
limit.

To aid searching for other white dwarf subtypes beyond the DAs and DBs, we selected all objects 
that had not been successfully fitted as a DA or DB star by
\autofit and then further selected only those with
$(g - r) < 0.5$, $(u - g) < 0.8$, and 
$g < 19.5$.  The color cuts helped to limit interlopers since most white
dwarf stars fall within these ranges and the magnitude cut simply helps
ensure a signal to noise ratio high enough to allow spectral
identification.  We thus obtained a list of 7864 objects that were then 
spectrally classified as discussed below.
In general, we looked for the following features to aid in the
classification for each specified white dwarf subtype:

\begin{itemize}
\item Balmer lines --- normally broad and with a Balmer decrement  
[DA but also DAB, DBA, DZA, and subdwarfs]
\item HeI 4471\AA\ [DB, subdwarfs]
\item HeII 4686\AA\ [DO, PG1159, subdwarfs]
\item C2 Swan band or atomic CI lines [DQ]
\item CaII H \& K  [DZ, DAZ]
\item CII 4367\AA\ [HotDQ]
\item Zeeman splitting [magnetic DA]
\item featureless spectrum with significant proper motion [DC]
\item flux increasing in the red [binary, most probably M companion]
\end{itemize}

Many of the stars analyzed in this way turned out to be genuine DA or DB
white dwarf stars
that had been rejected by \autofit for lack of signal to noise, too many bad
pixels in the spectra, uncertain colors, etc. Many were also
multi-subtype white dwarf stars like DAH, DBA, DAZ, DBZ or DA+M and DB+M.

We found many objects with both strong Balmer
lines and HeI lines. These objects are likely double
degenerate binaries composed of a DA and a DB white dwarf, but following
standard nomenclature, we simply classified them as DAB or DBA, as
appropriate. Another
group of objects had Balmer lines less deep than what is expected
for DA white dwarf stars with their derived effective temperatures.
These stars are also most likely
double degenerates consisting of a DA and a DC white dwarf stars, but were
classified as DA.
\citet{tre11} analyze these potential double degenerates. 

We also found a group of stars to have a very
steep Balmer decrement (i.e. only a broad H$\alpha$ and sometimes
H$\beta$ is observed while the other lines are absent) that could not
be fit with a pure hydrogen grid. We find that these objects are
best explained as helium-rich DAs, as confirmed by fits with a grid
of helium-rich white dwarf stars with traces of hydrogen 
\citep[see][]{duf07}.
These white dwarf stars are most probably former DZA where
all the metals have gravitationally settled while the hydrogen still
floats at the surface. 

For the other spectral types (DC, DZ, DQ and HotDQ), we used an
appropriate grid and fitted the spectroscopic and photometric data
\citep[see][for details]{duf05,duf07,duf07a,koe06a,duf08} to confirm the
classification. Objects that could not be fitted using one of the
grids were thus easily spotted and put in the non-white-dwarf
category.

We finally note that the white dwarf color space also contains
many hot subdwarfs. It is difficult, just by looking at a spectrum,
to tell a low mass white dwarf from a subdwarf. To guide us in the
classification of hot stars, we superposed over the observed spectra a
$\log{g}=7$ model at the effective temperature given by a fit to the
ugriz colors.  We then rejected objects showing lines much less broad than the
synthetic spectrum.   We also declared a subdwarf to be anything labeled as a
subdwarf (``SD''), DB, or DA by our \autofit code with a $\log{g}$ of 6.5
or less and human-classified as either a subdwarf or a DA or DB.  Since
the {\it autofit}-measured values of $T_\mathrm{eff}$ and $\log{g}$ are not reliable
for anything other than a clean DA/DB spectrum, anything
that passed the \autofit subdwarf criteria, but was human-classified as
anything other than a subdwarf or clean DA/DB was not classified as a
subdwarf. So, for example, some DAMs in our catalog that have low \autofit 
$\log{g}$ values may actually be subdwarf + M systems.

\subsection{Classification Results}
Table~\ref{tb:ids} lists the number of each type of white dwarf star we
identified.  Table~\ref{tb:columns} lists the columns of data provided
for in our electronic catalog file (in comma separated variable format).
The full catalog is available
in the online electronic version of this article.

\begin{deluxetable}{rrl}
\tablecolumns{3}
\tablewidth{0pc}
\tablecaption{\label{tb:ids}Numbers of identified white dwarf types.}
\tablehead{
\colhead{No. of Stars} & \colhead{No. of Spectra} & {Type}\\}
\startdata
12\,843 & 14\,120 & DAs\\
923 & 1011 & DBs\\
628 & 681 & DAH\\
10 & 22 & DBH\\ 
91 & 101 & Other Magnetic\\ 
559 & 605 & DC\\ 
409 & 447 & DZ\\ 
220 & 243 & DQ\\ 
61 & 65 & DO/PG~1159\\ 
1735 & 1951 & WD+MS\tablenotemark{a}\\ 
124 & 141 & WDmix\tablenotemark{b}\\ 
951 & 1020 & WDunc\tablenotemark{c}\\ 
\enddata
\tablenotetext{a}{These spectra show both a white dwarf star and a
companion, non-white dwarf spectrum, usually a main sequence M star.}
\tablenotetext{b}{These stars are mixed white dwarf subtypes. We did not attempt to resolve if the observed features resulted
from single star or multiple star systems.}
\tablenotetext{c}{These spectra were identified as uncertain white dwarf stars.}
\end{deluxetable}

\begin{deluxetable}{cll}
\tablecolumns{3}
\tablewidth{0pc}
\tablecaption{\label{tb:columns}Columns provided in data tables.}
\tablehead{
\colhead{Column No.} & \colhead{Heading} & {Description}}
\startdata
1 & Name & SDSS object name (SDSS 2000J+) \\
2 & Plate & SDSS plate number \\
3 & MJD & SDSS Modified Julian date \\
4 & Fiber & SDSS FiberID \\
5 & RA & Right ascension \\
6 & Dec & Declination \\
7 & SN\_g & SDSS g band signal to noise ratio \\
8 & u\_psf & SDSS u band PSF magnitude \\
9 & u\_err & SDSS u band PSF magnitude uncertainty \\
10 & u\_flag & SDSS u band quality control flag \tablenotemark{a}\\
11 & g\_psf & SDSS g band PSF magnitude \\
12 & g\_err & SDSS g band PSF magnitude uncertainty \\
13 & g\_flag & SDSS g band quality control flag \tablenotemark{a}\\
14 & r\_psf & SDSS r band PSF magnitude \\
15 & r\_err & SDSS r band PSF magnitude uncertainty \\
16 & r\_flag & SDSS r band quality control flag \tablenotemark{a}\\
17 & i\_psf & SDSS i band PSF magnitude \\
18 & i\_err & SDSS i band PSF magnitude uncertainty \\
19 & i\_flag & SDSS i band quality control flag \tablenotemark{a}\\
20 & z\_psf & SDSS z band PSF magnitude \\
21 & z\_err & SDSS z band PSF magnitude uncertainty \\
22 & z\_flag & SDSS z band quality control flag \tablenotemark{a}\\
23 & PM & SDSS proper motion (0.01 arcsec/yr) \\
24 & PM\_angle & SDSS proper motion angle (+North through East)\\
25 & PM\_match & SDSS proper motion match (1=successful match within 1.0
arcsec)\\
26 & A\_g & SDSS g band extinction \\
27 & GMT & SDSS mjd\_r (GMT when row 0 of r measurement read)\\
28 & AutoType & \autofit ID \\
29 & T\_eff & \autofit $T_{eff}$ \\
30 & T\_err & \autofit $T_{eff}$ uncertainty \\
31 & log\_g & \autofit $\log{g}$ \\
32 & log\_gerr & \autofit $\log{g}$ uncertainty\\
33 & chisq & \autofit $\chi^2$ fit measurement\\
34 & uniq & {\it unique} number \tablenotemark{b}\\
35 & Mass & calculated mass for clean DAs and DBs only\\
36 & Mass\_err & mass uncertainty\\
37 & humanID & human ID assigned this spectrum\\
\enddata
\tablenotetext{a}{The photometric flag values are processed versions of the
{\rm $flags_{\it band}$} parameter in the SDSS database.  It has been
logically anded with the appropriate values to highlight objects that
have the following quality control flags set: EDGE, PEAKCENTER, NOPROFILE,
BAD\_COUNTS\_ERROR, INTERP\_CENTER, DEBLEND\_NOPEAK, PSF\_FLUX\_INTERP,
SATURATED, and NOTCHECKED.  If the value is non-zero, then the corresponding 
SDSS magnitude is suspect.}
\tablenotetext{b}{The {\it unique} number is assigned to identify duplicate
spectra of the same object. For objects with only one spectrum, the value
of this column is {\it uniq}. For objects with more than one spectrum, the
value will be {\it dup-xxxx} where xxxx is a running number, the same for
all spectra of the same object.  The spectrum with the highest signal to noise
ratio for objects with duplicate spectra will be identified with an {\it
a} at the end of its {\it dup-xxxx} name.}
\end{deluxetable}

\section{Sample completeness}

As discussed in the \citet{kle04} and \citet{eis06} catalogs, the
spectroscopic sample of white dwarf stars from the SDSS is not at all
complete. That is, not every white dwarf star in the SDSS photometric 
survey has a corresponding SDSS spectrum.
There are many complex biases and selection effects based on the
many different criteria used to obtain these white dwarf spectra.
Each individual criterion has a different priority when it comes to
designating objects in a given part of the sky for follow-up spectroscopy,
Since the selection
criteria, weights, and biases, however, are all known,  the completeness
of our spectroscopic white dwarf sample is knowable, but is beyond the focus
of this work. 

Since we did not look at, or even fit, every SDSS spectrum in DR7, we do
need to estimate how complete our catalog is compared to the overall, but
unknown, SDSS spectroscopic white
dwarf sample. Did we recover every white dwarf with a spectrum in DR7?  One
way we can address this question is to see how many of the \citet{kle04} and
\citet{eis06} white dwarf stars we recovered in our new catalog since
our candidate selection and fitting
were done independently of the earlier catalogs.

When comparing our catalog with Table~11 from \citet{eis06}, we found 241
missing
out of 10\,090 objects or about 2.4\%.  We then added the missing
spectra to our candidate list and proceeded to analyze them along with the
rest of our candidates.  A spot check of these objects showed that they
don't pass the color cuts described in section~3.1 of \citet{eis06} in
either the SDSS DR4 or DR7 database. 
\citet{eis06} used some preliminary photometry in the
early candidate selection process and did not redo the candidate selection
once the released DR4 photometry became available, so this likely explains
why we did not pick them  up in our candidate list.
The completeness of this catalog, therefore, ought to be at least as good as
that  of \citet{eis06}.

Doing a similar test with Table~5 from \citet{kle04} resulted in
258 white dwarf stars  from the DR1 catalog
not making our candidate list.  53 of them belong to Plate-MJD pairs which
are not part of DR7 and 72 stars were the same as those
missing from the \citet{eis06} DR4 catalog.   So, combined, there were 183
out of 2971 stars from the DR1 catalog, or $\approx 6\%$  not included 
in either our original candidate list or in the DR4 catalog.  The candidate
selection parameter space in the DR1 catalog \citep{kle04}, though, was
much more extensive that that used in the DR4 catalog \citep{eis06}, and
thus, here. So that missing parameter space, if included here, would
probably result in another 600 or so white dwarf stars, most of which would
be cooler and overlap the A and F main sequence star region.

\subsection{Literature comparison}

\begin{figure}
\plotone{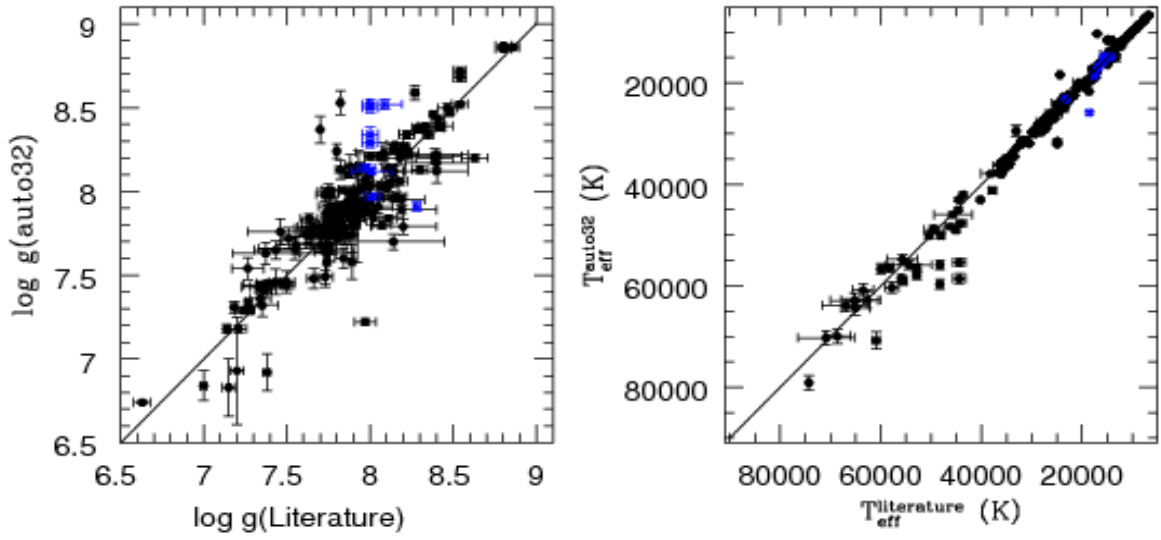}
\caption{Comparisons of our \autofit version {\it auto32} fits to atmospheric values in the
literature. The diagonal line shows the one to one correspondence.  The
black circles are DAs and the blue squares, DBs.
The larger scatter seen above $T_\mathrm{eff}\approx 50\,000K$ is due to our models
assuming local thermodynamic equilibrium (LTE) whereas non-LTE (NLTE) models 
are needed for these temperatures.\label{fig:literature}}
\end{figure}

Figure ~\ref{fig:literature} shows 
the comparison for 195 DAs and 10 DBs between our \autofit
determinations and those in the literature
(Oke, Weidemann \& Koester 1984,
Finley, Koester \& Basri 1997,
Friedrich et al. 2000,
Claver et al. 2001,
Gianninas, Bergeron \& Fontaine 2005,
Liebert, Bergeron \& Holberg 2005,
Lisker et al. 2005,
Kawka \& Vennes 2006,
Kepler et al. 2006,
Voss et al. 2007,
Kilic et al. 2007,
Lajoie \& Bergeron 2007,
Stroeer et al. 2007,
Holberg, Bergeron \& Gianninas 2008,
G{\"a}nsicke et al. 2008,
Nebot G\'omez-Mor\'an et al. 2009,
Pyrzas et al. 2009,
Allende Prieto et al. 2009,
Kilic et al. 2010,
Girven et al. 2010,
Kulkarni \& Kerkwijk 2010). 
We have not included in the comparison the values from DR1 or DR4.  The
figure clearly shows that our fit parameters are generally in very good
agreement with those from the literature, where overlap does occur.

\begin{figure}
\plotone{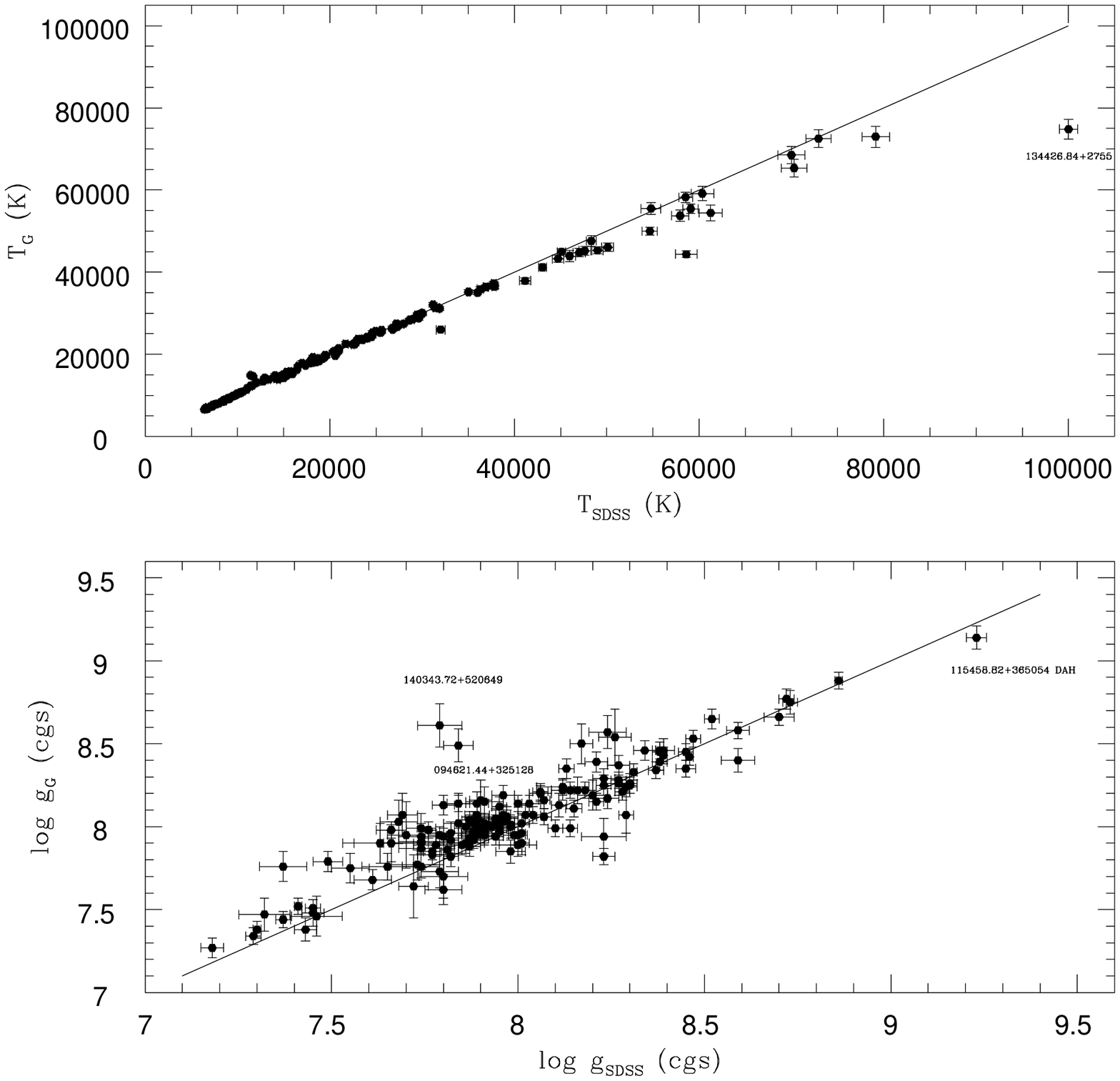}
\caption{Comparisons of our \autofit fits to atmospheric values in \citet{gia11}.}
\label{fig:gian}
\end{figure}

\citet{gia11} obtained $S/N\geq 50$ spectra of 177 white dwarfs in common with our sample
and estimated $T_\mathrm{eff}$ and $\log g$ values for those using the Montreal
group atmospheres with ML2/$\alpha=0.8$ convection theory, fitting the line profiles
only. In Figure~\ref{fig:gian}
we plot their determinations and ours for the 162 white dwarfs that don't show
a companion in our spectra. The agreement is good in spite of
the different spectra, different models, and different fitting procedures.
For stars hotter than $T_\mathrm{eff} \approx 50\,000K$, the differences are larger
because our models assume local thermodynamical equilibrium (LTE).

\subsection{Consistency}

\begin{figure}
\includegraphics[scale=0.65,angle=-90]{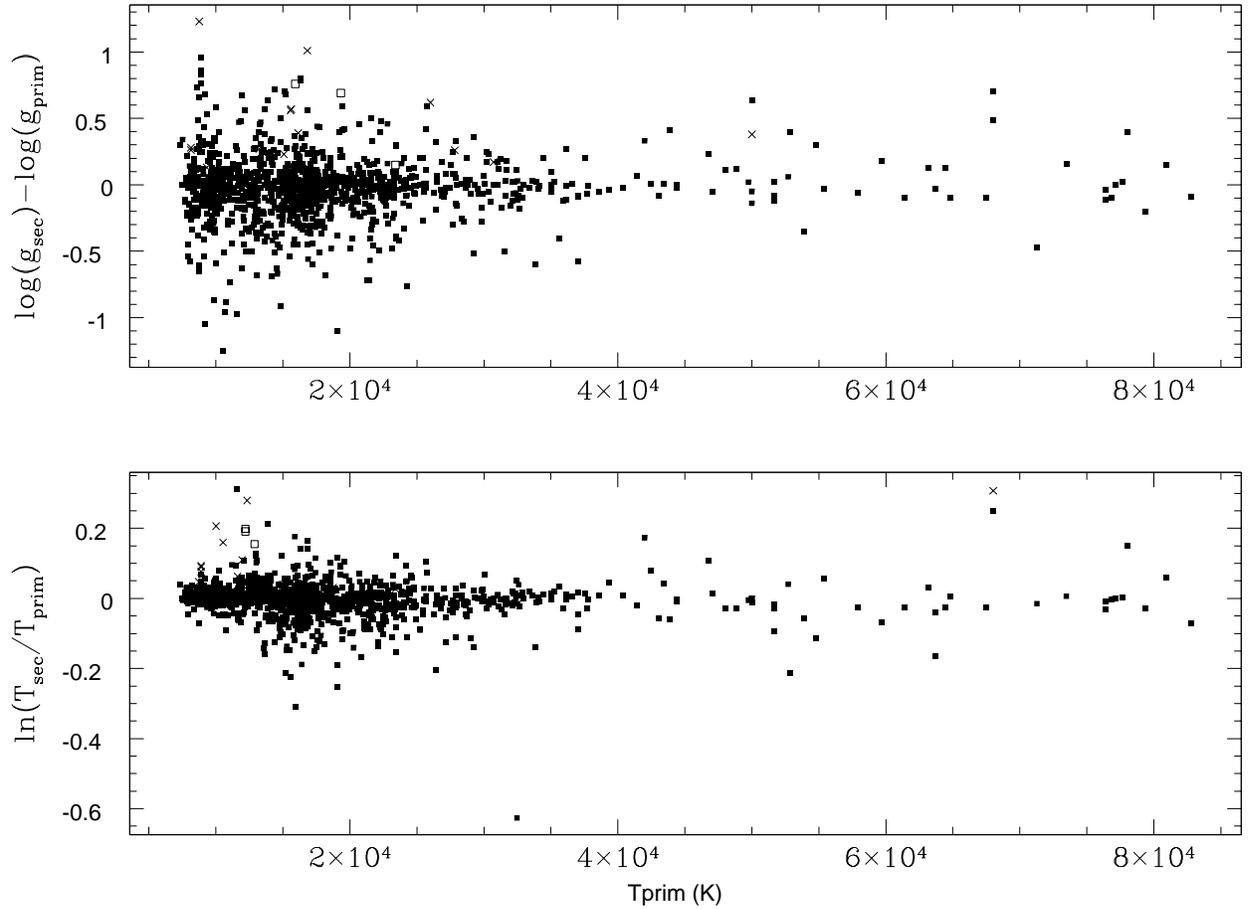}
\caption{\autofit comparisons of fits to duplicate spectra of clean DA and
DB white dwarf stars in our catalog. Solid squares represent objects where
the duplicate measurements differ by less than $3 \sigma$, the x's, between
3 $\sigma$ and 5 $\sigma$, and the hollow squares, $> 5 \sigma$. \label{fig:auto_dups}}
\end{figure}

Before rationalizing our identifications and removing the subdwarf spectra, 
we found 1683 objects in our catalog with more than one spectrum in
DR7.  There are a total of 3591 spectra for these 1683 objects, so most
duplicates have only two spectra in total.  Two objects have 7 DR7 spectra,
the most of all the duplicates.
Each 
duplicate spectrum was independently fit by \autofit and by eye so we
could use the duplicate IDs as a consistency check to our results.
Figure~\ref{fig:auto_dups} shows the resulting comparison from our \autofit
results.  The average absolute value of the difference in $T_\mathrm{eff}$
is 680K and the average abssolute value of the difference in $\log g$ is 0.16.

Of the set of 3591 duplicate spectra, only 242 of them had human
identifications that disagreed with each other.  212 of these agreed in the
dominant subtype with 13 differing only by our uncertainty note, a ":"
indicating an uncertain identification of the indicated
spectral feature.  141 identifications differed by an additional
subtype or a subtype and a ":", and 58 differed by more than one
subtype.  The remaining 30 identifications, $\approx 0.8\%$ of the sample,
disagreed in dominant subtype
and are mostly DAB/DBA, DA/DC, and DA/SDB pairs.  We examined each one
of these disagreeing IDs and selected the best identification (usually 
that of the highest signal to noise spectrum) and applied
it to all spectra for each object.
In general, we found our classifications were different only for low signal
to noise spectra.

We also compared our identifications with those made in the \citet{eis06}
DR4 catalog.  Of the 10\,090 WDs in the DR4 catalog, 8527 of our IDs agreed.
1563 of them disagreed.
Of these 1563
disagreements, 1330 agreed in dominant subtype, with 254 differing only by
a ":", 591 by an additional subtype or a subtype and a ":", and 485
by more than one subtype.  227, $\approx 2\%$, disagree in dominant subtype.  
\citet{eis06}
did not hand-identify each object in the catalog, relying on \autofit to
accurately report clean DA and DBs and to identify which subset of objects
to look at individually. Thus, this level of disagreement between our two
catalogs seems consistent with our 100\% hand-checked identifications.

\begin{figure}
\includegraphics[scale=0.70]{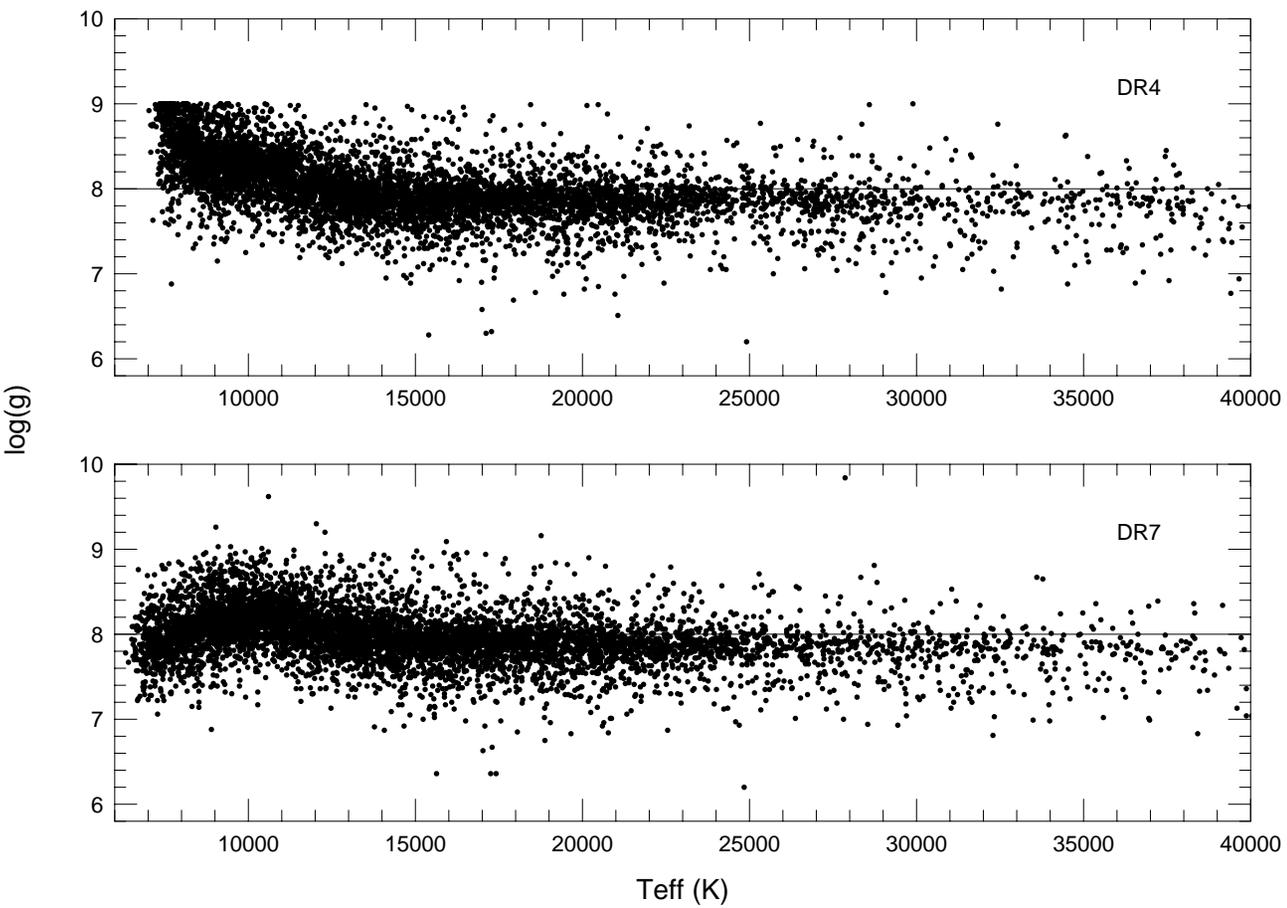}
\caption{A comparison of DA $\log{g}$ vs. $T_\mathrm{eff}$ \autofit values for 
DR4 stars also in DR7. The top panel shows the DR4 values while the bottom
panels shows our new determinations. Our improved model physics have reduced 
the rise to higher $\log{g}$ at lower temperatures to a bump, improving, but 
not completely eliminating this well known model artifact \citep[eg.,][]{tre10}.}
\label{fig:dr4-dr7DAAutofit}
\end{figure}

\begin{figure}
\plotone{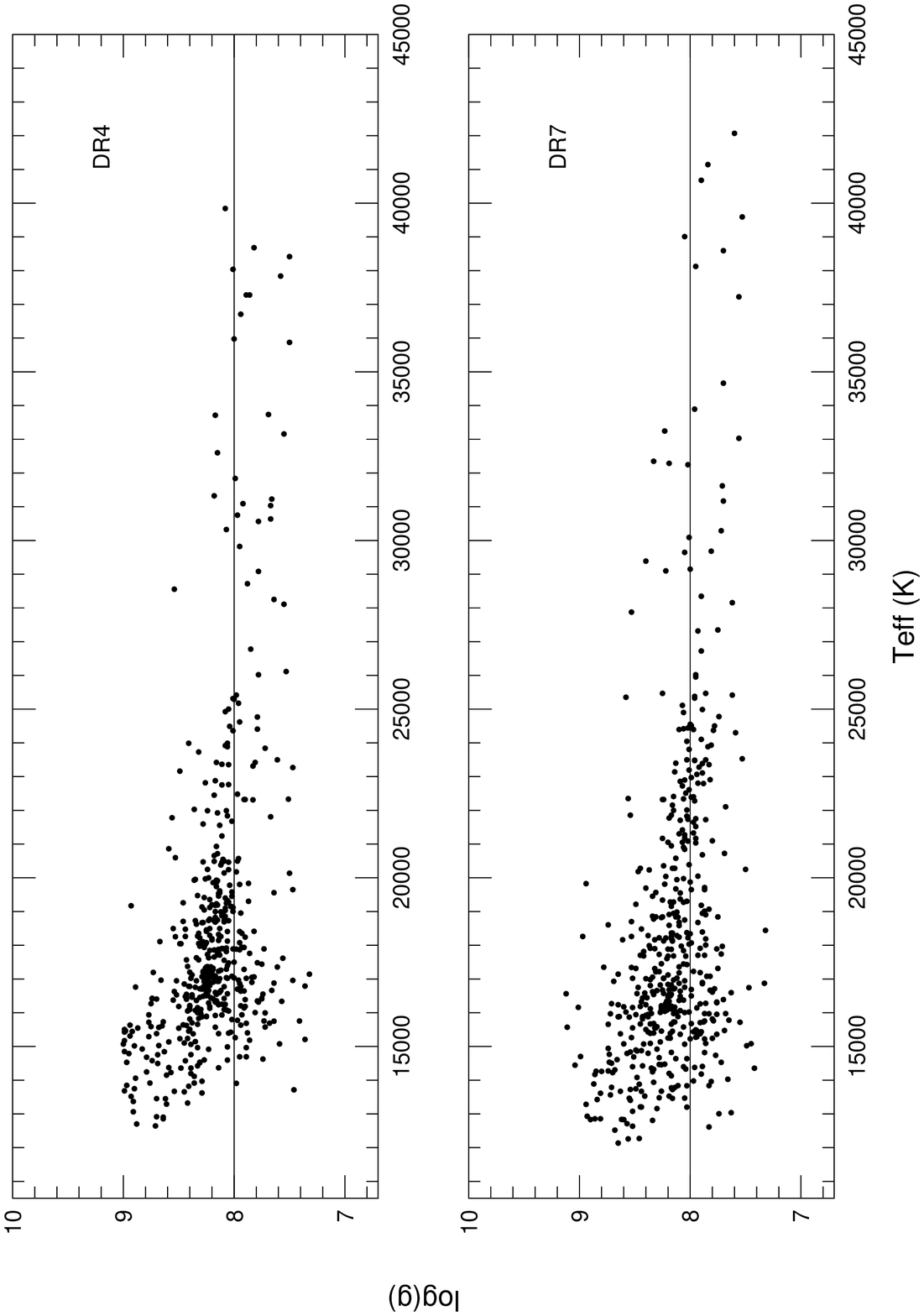}
\caption{A comparison of DB $\log{g}$ vs. $T_\mathrm{eff}$ \autofit values for 
DR4 stars also in DR7. The top panel shows the DR4 values while the bottom  
panels shows our new determinations.}
\label{fig:dr4-dr7DBAutofit}
\end{figure}

Comparisons of our \autofit parameters of the DR4 stars in DR7 with the DR4
fits is a measure of the changes to both our \autofit models and the DR7
SDSS spectral reductions.
Figures~\ref{fig:dr4-dr7DAAutofit} and~\ref{fig:dr4-dr7DBAutofit} show the
DA and DB $\log{g}$ and $T_\mathrm{eff}$ distributions for the DR7 objects in
DR4.  The biggest changes occur at the cool end of the DA distribution where
the DR4 fit values rise to higher $\log{g}$ with lower $T_\mathrm{eff}$, a change due to
the improved input model neutral broadening physics discussed earlier.
Our new model grid's increase in range in $\log{g}$ and $T_\mathrm{eff}$ is
also clearly evident in these figures.

\section{Results}
Besides producing the catalog itself, which we hope will spawn a lot of
future papers and analysis, we report here on the increased number of
magnetic white dwarf stars found in this catalog, compared with those
previous. We also look at the mass distribution of our DA and DB samples
and find a decidedly non-Gaussian DA mass distribution and a statistically
significant difference in mean mass between the DAs and the DBs.  As in
\citet{eis06} and \citet{eis06a}, we again find no statistically
significant DB gap.

\subsection{Magnetic Fields and Zeeman Splittings}
When examining each candidate spectrum,
we found hundreds of stars with Zeeman splittings
indicating magnetic fields above 
3~MG (the limit below which we do not think we can accurately identify) that if
not identified as magnetic in origin, would have rendered 
inaccurate \autofit $T_\mathrm{eff}$ and 
too high $\log g$ determinations.  We ended up classifying
628 DAHs, 10 DBHs and 91 mixed atmosphere magnetics, compared to
only 60 magnetic white dwarf stars of all types identified in
\citet{eis06}.  

\citet{sch03} found 53 magnetic white dwarf stars in DR1 and \citet{van05} found 
52 in DR3
data.  Most of these stars did not make the \citet{eis06} DR4 catalog because they
did not meet the candidate selection criteria.

We also identified several hundred possible magnetic stars with
low signal to noise spectra that made solid identifications difficult.
These objects are accounted for in the {\em WDunc} category in Table~\ref{tb:ids}.
Wishing to not bias our mass distribution with a magnetic sample, we chose
to label them uncertain magnetic (DH:) stars since even fields as low as
$\approx 3$ or more MG will affect \autofit gravity determinations substantially.

\citet{kul09} independently found 44 of our
newly-classified
magnetic DAs and fitted the SDSS spectra to atmospheric models
including off-centered dipoles, assuming $\log{g}=8$. 

Here, we are reporting the nunmber of magnetic white dwarfs stars relative
to those non-magnetic to be roughly 3.5\%.
This number is in reasonable
agreement with \citet{sch95}, for example, but is significantly larger
than the $\approx 0.1\%$ and $1.5\%$ reported in the DR4 \citep{eis06}
and DR1 \citep{kle04} catalogs, respectively.  The DR4 catalog was based
primarily on computer identifications, supplemented by only partial human
checks, thereby explaining part of the reason for the lower numbers of
magnetic white dwarf identifications in \citet{eis06}.  Beyond this
cause, however, the algorithm we used to manually identify possible magnetic
white dwarf stars developed significantly since the DR1 and DR4 catalogs.
Combined with our desire to develop a clean sample for mass estimation, we
ended up with the larger, though plausible, percentage gain of magnetic
white dwarf stars seen here.

\begin{figure}
\plotone{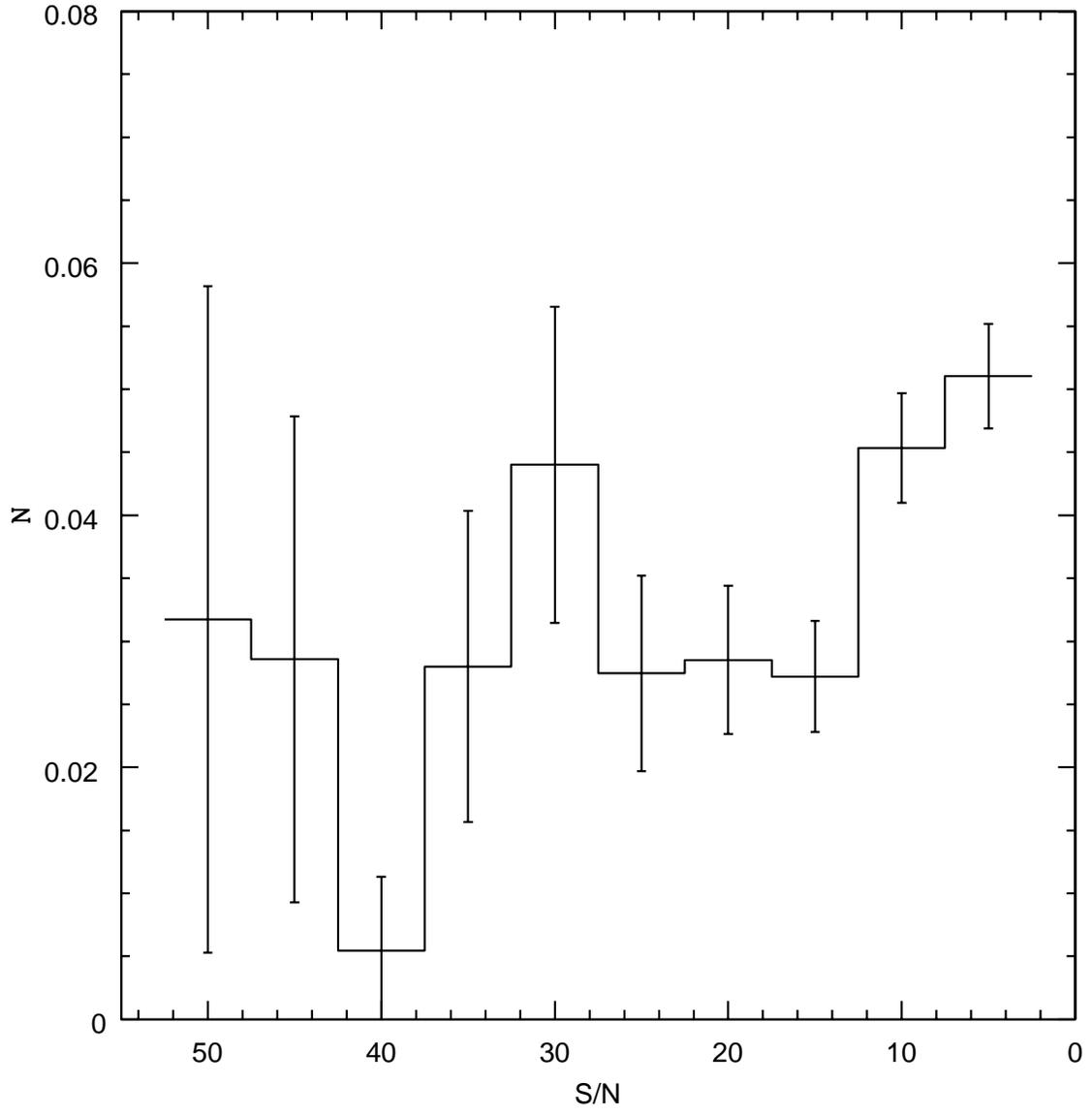}
\caption{Percentage of identified magnetic white dwarf spectra as a function
of spectral signal to noise.}
\label{fig:magvsn}
\end{figure}

\begin{figure}
\plotone{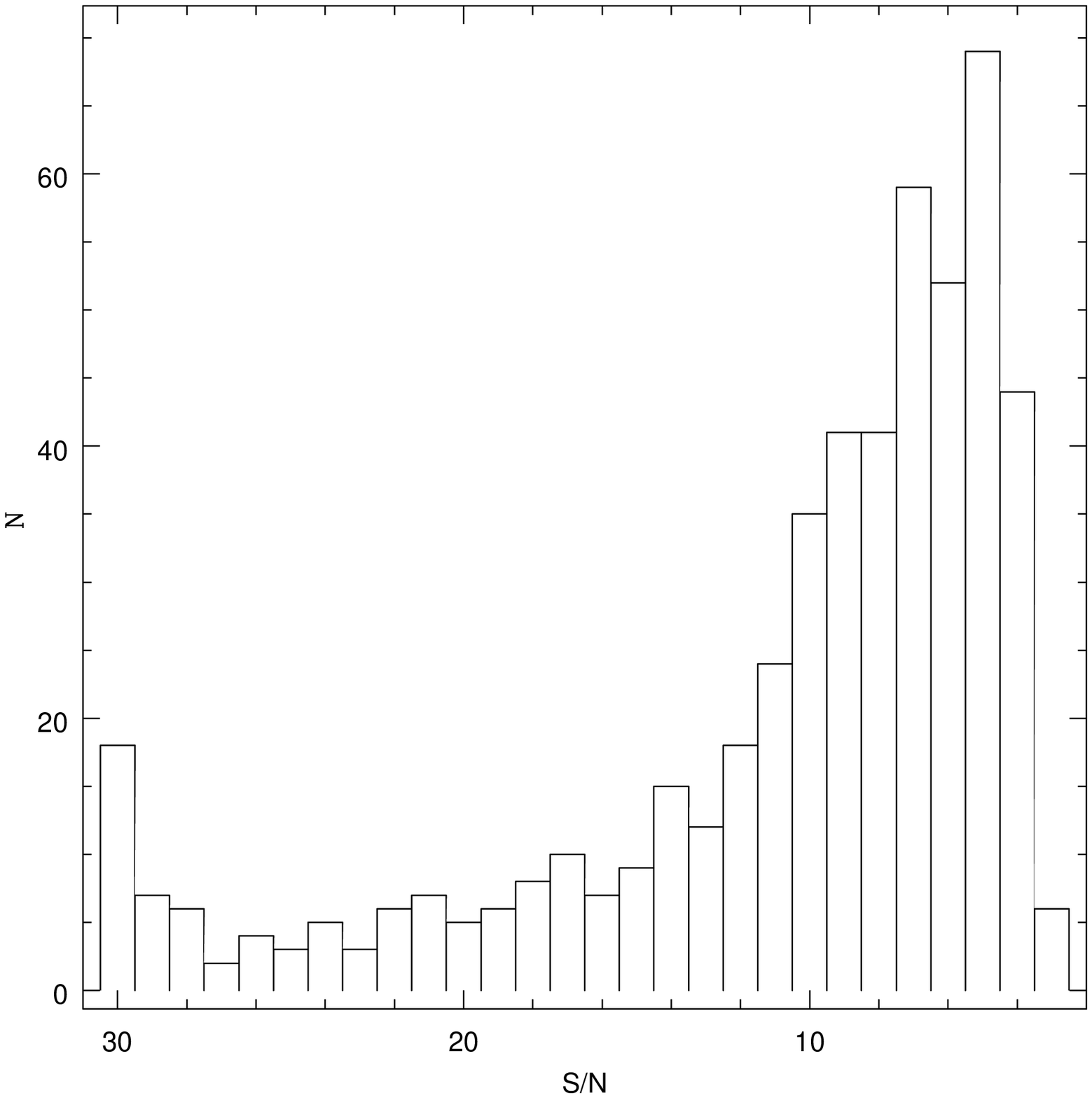}
\caption{The number of identified DAH stars as a function of spectral signal
to noise.}
\label{fig:dahsn}
\end{figure}

To validate our identification methods, we conducted several blind
simulations where we hand-identified model spectra with varying
amounts of added noise and magnetic field strength (from 0 to 800MG).
These simulations are reported in more detail in \citet{kep12}, but the
summary result is that our human identifications were proven valid for
spectra with $S/N >8$ and $B >2MG$.   In addition, at the very largest
magnetic field strengths, we found we are not identifying all the magnetic
white dwarf stars, instead labeling them as unknown stars or sometimes, DC.
Figure~\ref{fig:magvsn} shows the percentage of identified magnetic
DA white dwarf stars as a function of spectral signal to noise and
Figure~\ref{fig:dahsn} shows the absolute number of DAH identifications per
signal to noise bin.  These plots
suggests that below $S/N<8$, at least half of our $\approx 280$ DAH 
identifications are
likely magnetic and that spectra below $\approx S/N<8$ ought to be
confirmed with higher signal to noise and possibly higher resolution
spectra before declaring them magnetic or not.

\subsection{Mean masses}

To calculate the the mass of our identified clean DA and DB stars
from the $T_{\mathrm{eff}}$ and $\log g$ values
obtained from our fits, we used 
the mass-radius relations of
\citet{ren10} for
carbon-oxygen DA  white dwarfs.  These  relations are
based on  full evolutionary calculations appropriate for  the study of
hydrogen-rich  DA  white  dwarfs  that  take  into  account  the  full
evolution of progenitor stars from the zero-age main sequence through
the  core hydrogen-burning  phase, the  helium-burning phase,  and the
thermally  pulsing asymptotic  giant branch  phase.  The  stellar mass
values  of the resulting  sequences are:  0.525, 0.547,  0.570, 0.593,
0.609,  0.632,  0.659,  0.705,  0.767, 0.837,  and  0.878~$M_\odot$.  These
sequences  are  supplemented  by  sequences  of  0.935  and  0.98~$M_\odot$
calculated specifically for this  work. For high-gravity white dwarf stars,
we employed the mass-radius relations for oxygen/neon core white dwarf
stars given
in \citet{alt05} in the mass range from 1.06 to 1.36~$M_\odot$ with
a step  of 0.02~$M_\odot$.  For  the low-gravity white dwarf stars,  we used the
evolutionary   calculations  of   \citet{alt09a} for
helium-core white dwarf stars.  These sequences are characterized by stellar
mass  values of  0.22, 0.25,  0.303, 0.40  and 0.452~$M_\odot$.  They were
complemented with  the sequences  of 0.169 and  0.196~$M_\odot$  taken from
\citet{alt10}.

For DB  white dwarf stars, we  relied on the evolutionary
calculations  of  hydrogen-deficient  white  dwarf stars of  0.515,  0.530,
0.542, 0.565,  0.584, 0.609, 0.664,  0.741 and 0.870~$M_\odot$  computed by
\citet{alt09b}. These  sequences constitute  an improvement
over previous calculations. In particular, they  have been derived from
the born-again
episode responsible for the  hydrogen deficiency.  For high-gravity DBs,
we used  the oxygen/neon evolutionary sequences described above
for the case of a hydrogen-deficient composition.

These evolutionary sequences constitute a complete and homogeneous grid
of white dwarf models that captures the core features of progenitor
evolution, in particular the internal chemical structures expected in the
different types of white dwarf stars. 

To calculate reliable mass distributions, we selected only the best S/N
spectra with temperatures well fit by our models.  We find reliable
classifications can be had from spectra with S/N $\geq$ 15,
in agreement with \citet{tre11}.
We classified 14\,120
spectra as clean DAs, but selecting the highest S/N
spectra for those with
duplicate spectra, we are left with 12\,813 clean DA stars.
Of these DAs, 3577 have a spectrum with S/N $\geq 15$, with a mean S/N=$25\pm 10$. Using this sample, we obtain $\langle M_\mathrm{DA}
\rangle = 0.623 \pm 0.002 M_\odot$.

This mean mass estimate is incorrect, however, if we believe the apparent
increase in fit $\log{g}$ at low $T_{eff}$ is an artifact of our models and
not inherent in the stars.
Although \citet{lie03},
\citet{kep07}, \citet{kep09}, \citet{tre11}, and \citet{gia11} all show an 
increase in measured $\log{g}$ for DAs with
measured temperatures of order 12\,000--13\,000K or less,
this increase is probably due to missing physics in
the models and not due to the stars, since the photometric
determinations  and gravitational redshifts \citep{koe06,fal10}
do not show this $\log{g}$ increase.

We therefore further restricted our sample to those
DAs with a measured $T_\mathrm{eff}>13\,000K$, and from a sample of now
2217 objects, we determined
$\langle M_\mathrm{DA}^{T_\mathrm{eff}>13\,000~K} \rangle = 0.593 \pm 0.002 M_\odot$.

Our mean DA mass is smaller than that of \citet{tre09}, even though
we are using the same Stark broadening and microfield as they are.
Our sample is 5 times larger, however, and we have removed suspected
magnetic DAs from our sample, which would otherwise increase our measured
mean mass.
\citet{lim10},
with the same models as \citet{tre09}, obtained a mean mass
of $0.606 \pm 0.135~M_\odot$ for their KISO survey DA sample,
and $0.758 \pm 0.192~M_\odot$ for their DB sample.
\citet{fal10} determined the mean ensemble mass of a sample of
449 DAs observed in the ESO SN Ia progenitor survey (SPY) project, 
using their mean gravitational redshift
and found $\langle M \rangle=0.647\pm 0.014~M_\odot$, a value substantially
higher than ours. 
This value is independent
of the line profiles themselves and therefore should not be affected by linear
magnetic field effects. At the resolution of the SPY data \citep{koe01},
\citet{koe09a}
were able to identify fields larger than 90~kG, so the contribution of nonlinear
magnetic effects should also be negligible. 
\citet{gia11}, using $ML2/\alpha=0.8$ models, find 
$\langle M \rangle=0.638~M_\odot$ by fitting line profiles for high S/N spectra
for stars hotter than $T_\mathrm{eff}=13\,000$~K.  \citet{rom12} also
report a mean mass of $0.636~M_\odot$ for their asteroseismological
analysis of 44 bright pulsating DA (ZZ~Ceti, or DAV) stars.

If $\approx 10\%$ of our $S/N>15, T_\mathrm{eff}>13\,000K$ stars were
non-magnetic with masses of order $0.9~M_\odot$, our mean DA mass
would rise to $\approx 0.63~M_\odot$, more consistent with previous
measurements.  Such a number, though, would require all our identified
magnetic stars, and then some, to be both massive and non-magnetic, so
our lower mass estimate can be only partially explained by our magnetic
white dwarf identifications.  Additionally, we note that the \citet{tre11} 
re-analysis of the DR4 white dwarf stars resulted in
a measurement of $\langle M_\mathrm{DA} \rangle=0.613~M_\odot$, again lower
than most previous studies and similar to our new measurement.

\begin{figure}
\plotone{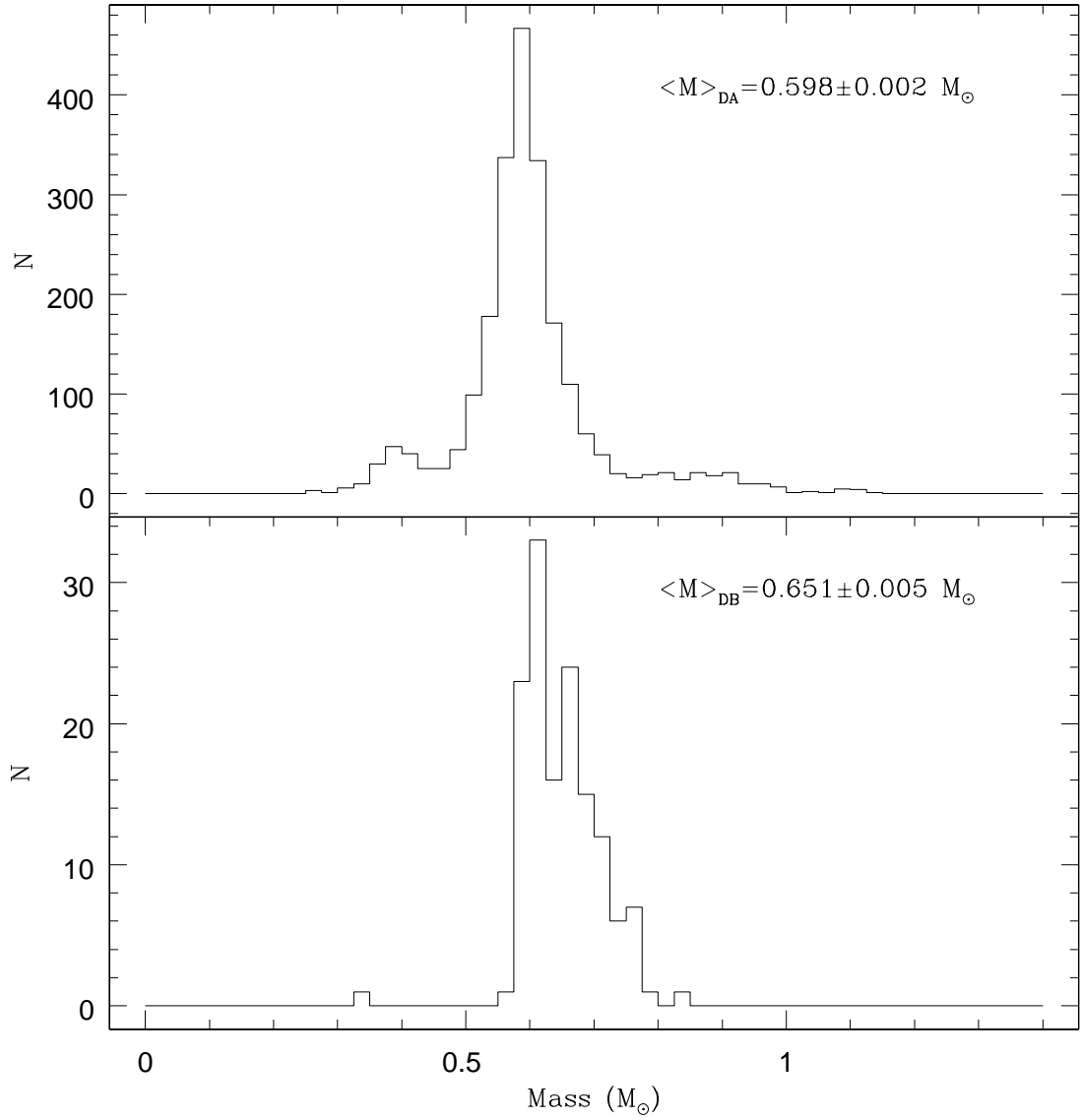}
\caption{Histogram of the masses for S/N$\geq 15$ clean DAs (upper panel) hotter than
$T_\mathrm{eff}=13\,000$~K and DBs (lower panel) hotter than
$T_\mathrm{eff}=16\,000$~K.
\label{fig:hist}}
\end{figure}

The 1011 spectra we classified as clean DBs belong to 923 stars. 
191 of these have a spectral S/N $\geq 15$, with a mean S/N=$23\pm 7$.  
Using this high S/N sample, we obtain
$\langle M_\mathrm{DB} \rangle = 0.685 \pm 0.013 M_\odot$.  Restricting
this sample to just those hotter than $T_\mathrm{eff}=16\,000$~K, again
assuming the increase in measured $\log{g}$ seen below this temperature is
an artifact of the models and not inherent to the stars,
we obtain 140 stars,
resulting in $\langle M_\mathrm{DB}^{T_\mathrm{eff}>16\,000~K}
\rangle = 0.676 \pm 0.014 M_\odot$. 
The masses of DAs and DBs are therefore statistically different,
as also found by 
\citet{kep07}, \citet{tre10} and \citet{ber11}.
This difference
is consistent with the possibility that DBs come through a very late thermal 
pulse phase, which
burns the remaining surface H after reaching the white dwarf cooling phase.
Figure~\ref{fig:hist} shows the obtained mass histograms for our final DA and
DB samples.

\subsection{DB Gap}
\citet{lie86} and \citet{lie87} show that of the total of 98 DBs known
then, none were
known in the temperature range $3\,0000 \leq
T_\mathrm{eff} \leq 45\,000$~K. They created the term {\it DB gap} for this
observed dearth
of helium dominated atmosphere white dwarfs in this temperature region. 
Subsequent DB studies did
not find more than one star in this temperature range, until the 10 to 28
DBs and DOs found in DR4 by \citet{eis06a}.
\citet{hug09} show that NLTE effects change the measured
$T_\mathrm{eff}$ by less than 15\% over those obtained by LTE atmosphere models
like we use, too small a difference to move all the stars within the
DB gap, outside of it.

Of the 923 stars we classified as clean DBs, we find 9
hotter than $T_\mathrm{eff}=45\,000$~K, 30 with $45\,000K \geq T_\mathrm{eff}
\geq 30\,000$~K, and 231 with $30\,000K \geq T_\mathrm{eff} \geq 20\,000$~K.
If we restrict ourselves to only the 57 (of the 923) stars with S/N$\geq 25$,
we find 1 hotter than $T_\mathrm{eff}=45\,000$~K,
 3 with $45\,000K \geq T_\mathrm{eff} \geq 30\,000$~K, and 18 with $30\,000K
 \geq T_\mathrm{eff} \geq 2\,0000$~K, following
the ratio expected from their ages 
\citep[1/6:1/18:1,][e.g.]{alt09b}.

Our numbers are in line with the \citet{eis06}
finding that there is a decrease in the number, although not a {\it gap},
 of DBs around
30\,000--45\,000K in relation to the hotter DO range.  

As evidenced in \citet{ber11}, for example, fitting a large sample such as
this with model DB stars consisting of pure Helium atmospheres, as we have 
done here, is not 
completely correct.  Even small (ie., undetectable) amounts of trace hydrogen
in the helium layer of a DB can cause significant errors to $T_{eff}$ and
$\log{g}$ determinations made by fitting pure helium outer atmospheres.
Hence, our results here are indicative of an avenue worth
exploring, but more detailed fitting may be needed to arrive at more
concrete conclusions.

\section{Conclusions and Discussion}
By classifying all our candidate white dwarf spectra by eye, supplemented
by our \autofit fitting of our DA and DB spectra, we roughly doubled the
number of previously known white dwarf stars and formed a large sample of
clean DA and DB spectra in order to study their mass distributions.
Our identifications are conservative in that we wanted to make sure we had
a clean DA and DB samples for our mass analyses, so we erred, if at all, on
the side of overinterpreting the spectra rather than underinterpreting them.
As a result, we identified a number of low field magnetic white dwarf stars
that represent a 5-fold increase in the number of known magnetic white dwarf stars.  We nonetheless 
believe these identifications are correct and suggest that previous
mass distribution analyses may have been biased towards higher masses,
given that these low-field magnetic stars were not previously recognized
as magnetic in earlier mass distribution measurements.

\begin{figure}
\plotone{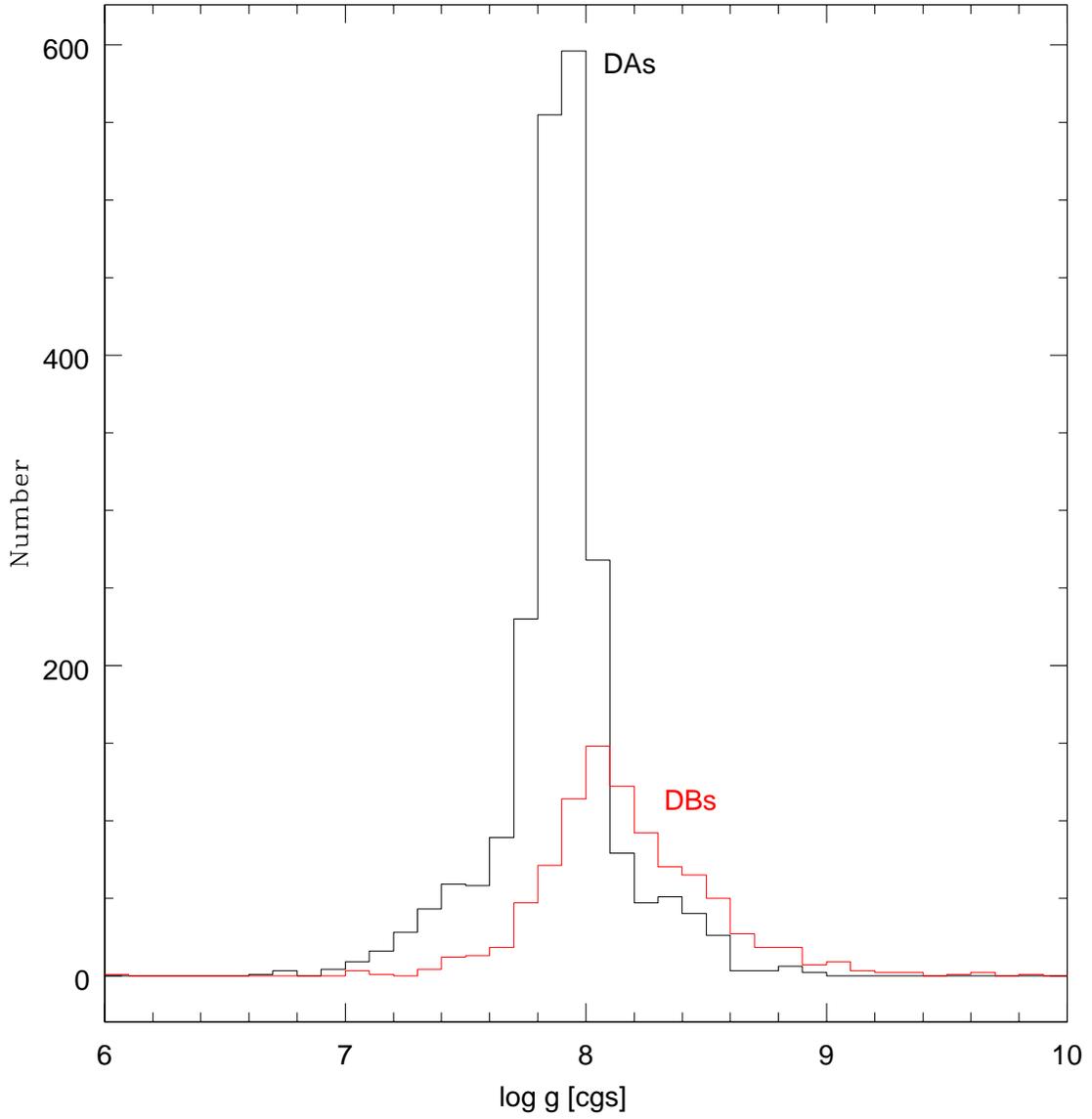}
\caption{Histogram of the surface gravity distribution of DAs (top black
unlabeled curve) and DBs (lower red labeled curve),
showing DBs first appearing at higher $\log g$ than DAs.
\label{fig:histg}}
\end{figure}

Our mean masses were determined only for stars with spectra conservatively 
identified as clean DAs and DBs with $S/N\geq 15$.
Perhaps as a result of this careful spectral selection, or perhaps as a
result of our increased sample size, we find
mean masses for DAs and DBs are smaller than those
by \citet{fal10} and \citet{gia11}.  Our comparisons to previous literature
determinations, where available, reveal no obvious biases in our
measurements, although the various magnitude and color limits
in the different targeting categories, and the signal to noise ratios required 
for accurate identifications are certainly selection
effects that could subtly affect our results.  Furthermore, as Figure~\ref{fig:histg} shows, our $\log{g}$
distributions qualitatively reproduce the difference we found in the mass
determinations, indicating the discrepancy is at least not solely due to
our conversions from $\log{g}$ to mass.

\begin{figure}
\plotone{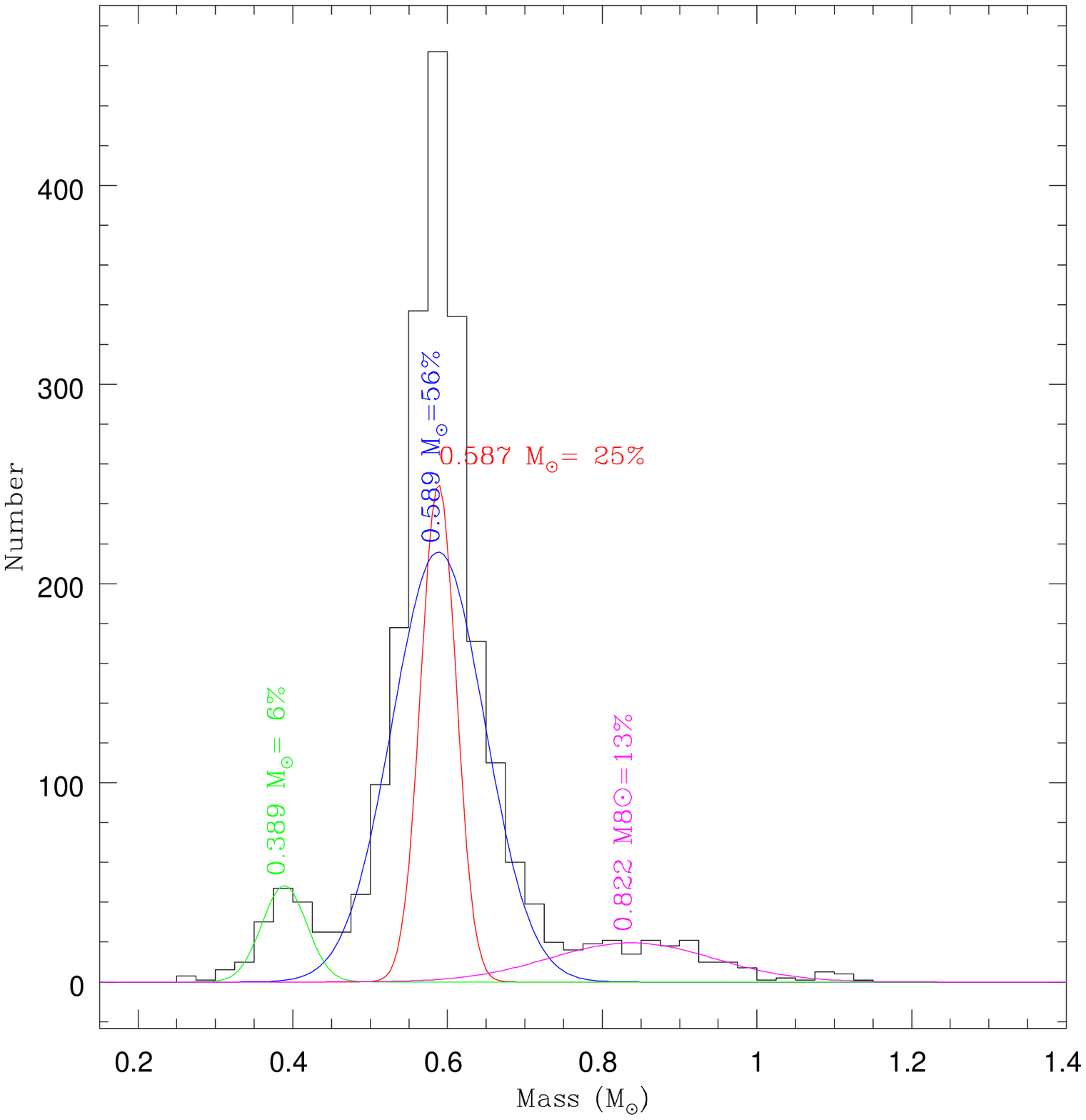}
\caption{Histogram of the masses for S/N$\geq 15$ clean DAs hotter than
$T_\mathrm{eff}=13\,000$~K and its Gaussian decomposition.
\label{fig:histe}}
\end{figure}

\begin{figure}
\plotone{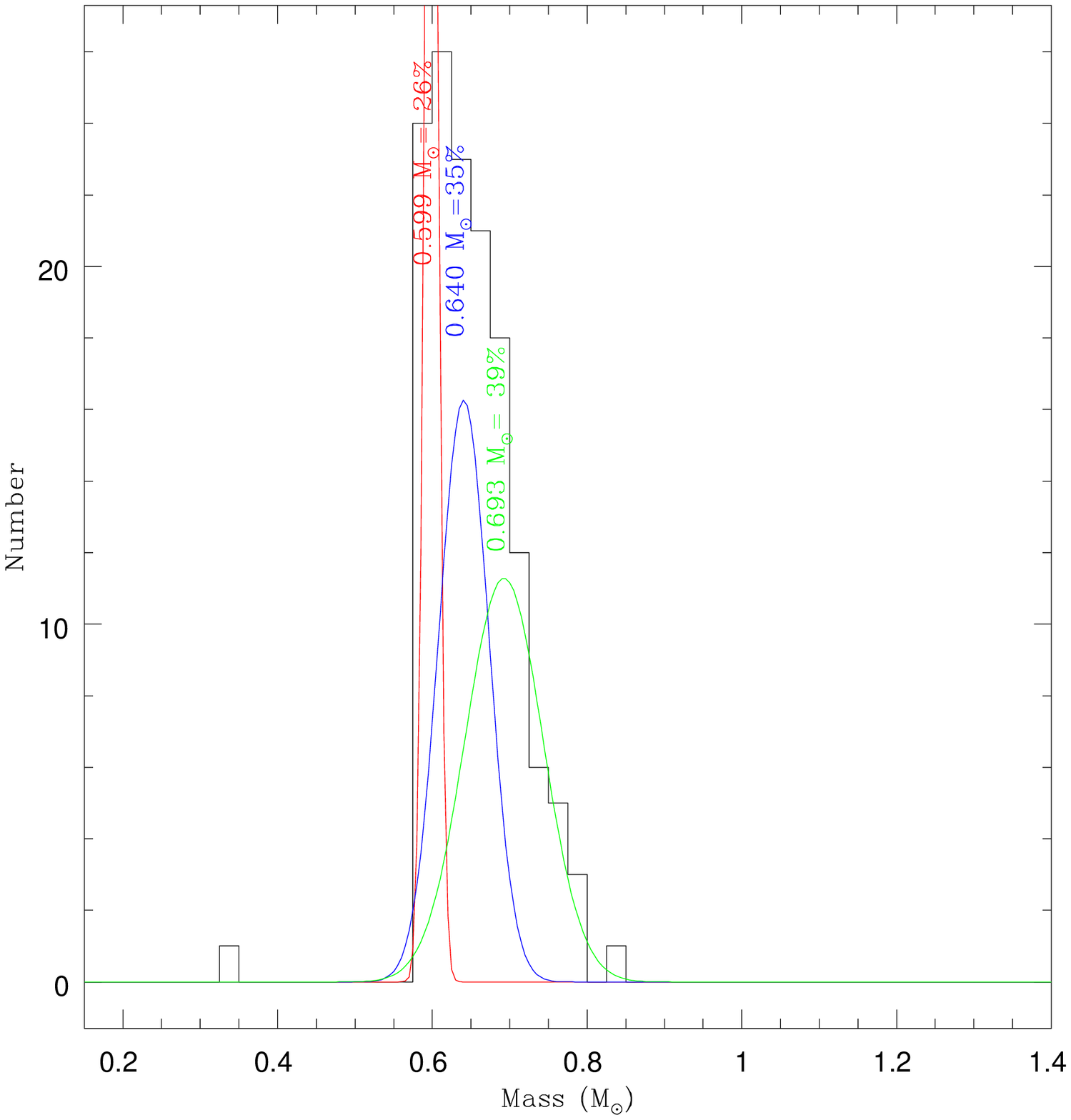}
\caption{Histogram of the masses for S/N$\geq 15$ clean DBs hotter than
$T_\mathrm{eff}=16\,000$~K and its Gaussian decomposition. }
\label{fig:histedb}
\end{figure}

There is no reason to expect the observed
mass distribution to be Gaussian. The ingredients are the
initial mass function, initial to final mass relation,
star formation rate,
and mass loss rates,
all of which are more or less well defined physical
non-Gaussian relations.
We find it informational, however, to use Gaussian deconvolutions of the 
mass distributions so we can talk about average/peak masses with some quantifiable
meaning attached.  We do not claim that each Gaussian component represents a
unique contribution to the DA/DB population.
Figures~\ref{fig:histe} and ~\ref{fig:histedb} show the DA and DB mass
distributions, respectively, broken down into their Gaussian components.
Table~\ref{tb:gfits} lists the mass peaks and percentage of objects
contained within each.  The figures clearly indicate that talking about a
mean DA mass is not particularly useful, but a peak mass (seen here at
$0.59 M_\odot$) is more useful.  The low and high mass wings of the DA
distributions are not symmetric, nor should they likely be, given the
age of the Universe ultimately determining the low mass cutoff for single
white dwarf stars.  The 
low mass DA component, at $0.43~M_\odot$ with 4\% of the
stars, is probably caused by binary interactions
since single star evolutionary models cannot generate these stars in
a Hubble time, while the smaller high mass peak at $0.82~M_\odot$ is likely
due to mergers.
The peak of the DB mass distribution (Figure~\ref{fig:histedb}) at 
$\approx 0.6M_\odot$ is similar that of the DA distribution although
the overall shape is quite different.

\begin{deluxetable}{lrc}
\tablecolumns{3}
\tablewidth{0pc}
\tablecaption{\label{tb:gfits}Gaussian components of observed DA and DB mass distributions.}
\tablehead{
\colhead{Component} & {Mean} & {Strength}\\
\colhead {} & ($M_\odot)$ & (\%) \\}
\startdata
DA & & \\
1 & 0.589 & 56 \\
2 & 0.587 & 25 \\
3 & 0.822 & 13 \\
4 & 0.389 & ~6 \\
DB & & \\
1 & 0.693 & 39 \\
2 & 0.640 & 35 \\
3 & 0.599 & 26 \\
\enddata
\end{deluxetable}

\begin{figure}
\plotone{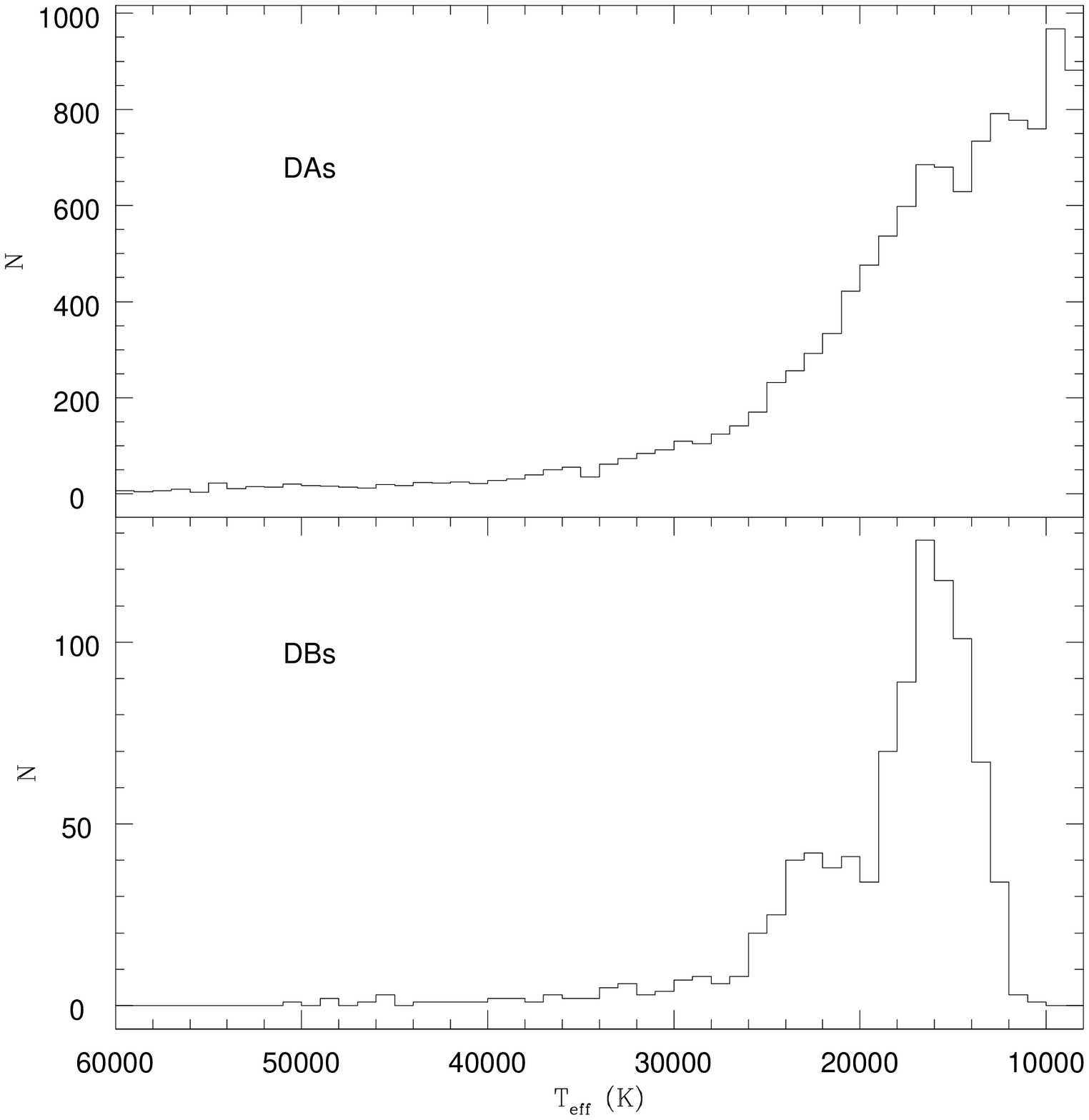}
\caption{Histogram of the $T_\mathrm{eff}$ distribution for our DA (top) and DB
(lower) samples.}
\label{fig:histteff}
\end{figure}

The $T_{eff}$ distributions shown in Figure~\ref{fig:histteff} do not
reveal a DB gap,
i.e., we do detect stars with HeI dominated atmospheres
hotter than $T_\mathrm{eff}=30\,000$~K, but there does seem
to be a decrease of cooler DBs, which simply become DCs, DQs, DZs
as they cool below $\approx 10\,000$K.

Although our measured mean mass for our DB sample is higher than that of our
DAs, the mass distributions show that the
largest number of DBs have masses similar to peak DA distribution.
The high mass tail of the DB distribution could be the result of a varying
number of thermal pulses or of varying metallicity in the DB progenitors.
It may also be that the higher mass DB progenitors are simply more prone to experience
a very late thermal pulse than are lower mass DB progenitors.
The trace low mass component in the the DB distribution may be 
associated with AM CVn stars, double He WDs, and a result of binary
evolution.

\acknowledgements
We thank Matt Burleigh for thorough and useful comments during the referee
process.

This work was partially supported by the Gemini Observatory which is operated 
by the Association
of Universities for Research in Astronomy, Inc., on behalf of the
international Gemini partnership of Argentina, Australia, Brazil, Canada,
Chile, the United Kingdom, and the United States of America.

Funding for the SDSS and SDSS-II has been provided by the
Alfred P. Sloan Foundation, the Participating Institutions, the
National Science Foundation, the U.S. Department of Energy, the
National Aeronautics and Space Administration, the Japanese
Monbukagakusho, the Max Planck Society, and the Higher Education
Funding Council for England. The SDSS Web Site is
http://www.sdss.org/.

The SDSS is managed by the Astrophysical Research Consortium
for the Participating Institutions. The Participating Institutions
are the American Museum of Natural History, Astrophysical Institute
Potsdam, University of Basel, University of Cambridge, Case Western
Reserve University, University of Chicago, Drexel University,
Fermilab, the Institute for Advanced Study, the Japan Participation
Group, Johns Hopkins University, the Joint Institute for Nuclear
Astrophysics, the Kavli Institute for Particle Astrophysics and
Cosmology, the Korean Scientist Group, the Chinese Academy of
Sciences (LAMOST), Los Alamos National Laboratory, the
Max-Planck-Institute for Astronomy (MPIA), the Max-Planck-Institute
for Astrophysics (MPA), New Mexico State University, Ohio State
University, University of Pittsburgh, University of Portsmouth,
Princeton University, the United States Naval Observatory, and the
University of Washington.

PD is a CRAQ postdoctoral fellow. This work was supported in part by 
NSERC Canada and FQRNT Qu\'ebec.

SOK, IP, VP, and JES were supported by FAPERGS and CNPq-Brazil.

\newpage
{\it Facilities:} {SDSS}.

\section{SQL Query Listings}
This is the SQL code used to reproduce the \citet{eis06} candidate
selection criteria.

\begin{lstlisting}
SELECT  
s.plate, s.mjd, s.fiberid,
p.ObjId, p.psfMag_u, p.psfMag_g, p.psfMag_r, p.psfMag_i, p.psfMag_z, 
p.ra, p.dec, s.z, u.propermotion, t.bsz, t.zbclass

FROM SpecObjAll s INNER JOIN PhotoObjAll p ON s.bestObjId=p.ObjID
LEFT OUTER JOIN sppParams t on s.specObjID=t.specObjID
LEFT OUTER JOIN USNO u on s.bestObjID=u.ObjID

WHERE p.psfMag_u<21.5

AND p.extinction_r<=0.6

AND
--either -2 <u-g < 0.833 - 0.667(g-r), -2 < g-r < 0.2
  ( ((p.psfMag_u-p.psfMag_g between -2 and
                                  (0.833-0.667*(p.psfMag_g-p.psfMag_r)))
     and (p.psfMag_g-p.psfMag_r between -2 and 0.2))
OR    -- or  0.2<g-r<1, |(r-i)-0.363(g-r)|>0.1,
    --    ( (u-g<0.7) or (u-g<2.5(g-r)-0.5)
  ((p.psfMag_g-p.psfMag_r between 0.2 and 1)
  and (Abs((p.psfMag_r-p.psfMag_i)-0.363*(p.psfMag_g-p.psfMag_r))>0.1)
  and ( ((p.psfMag_u-p.psfMag_g)<0.7)
     or ((p.psfMag_u-p.psfMag_g)<2.5*(p.psfMag_g-p.psfMag_r)-0.5) )))

AND
--the following is for specBS only; - low redshift and not a galaxy
(((t.bsz<0.003)
    and (t.zbclass <> 'GALAXY'))

OR
-- flags for all filters are OK 
((p.flags_u | p.flags_g | p.flags_r | p.flags_i | p.flags_z) &  
              (dbo.fPhotoFlags('INTERP_CENTER') |
               dbo.fPhotoFlags('COSMIC_RAY') |
               dbo.fPhotoFlags('EDGE') |
               dbo.fPhotoFlags('SATURATED'))=0)
AND -- small measured z with a good z determination
 (( (Abs(s.z)<0.003) and
    ((s.zwarning&1)=0))
OR -- large measured z, but with a high proper motion
 ( (Abs(s.z)>0.003) and
   (Abs(u.propermotion)>30) )))

ORDER BY s.plate, s.mjd, s.fiberid
\end{lstlisting}

The SQL code below was used to for the second half of candidate
generation as discussed in the text.

\begin{lstlisting}
-- find all objects targeted as WD, HOTSTD, 
-- classified as CV, CWD, WD, WDm,
SELECT t.plate, t.mjd, t.fiberID, 
       s.targettype, s.seguetargetclass, 
       s.sptypea, s.zbsubclass, 
       s.flag,
       t.primTarget, t.secTarget, t.seguePrimTarget, t.segueSecTarget
FROM sppParams s left outer join specObjAll t
ON 
     t.specobjid=s.specobjid
WHERE
 -- SEGUE TARGETing
     (s.targettype LIKE 'SEGUE_WD%') OR     
     (s.targettype = 'STAR_WHITE_DWARF') OR     
     (s.targettype = 'HOT_STD') OR     
     (s.seguetargetclass = 'HOT') OR
     (s.seguetargetclass = 'WD') OR
 -- SEGUE CLASSIFICATIONS
     (s.sptypea LIKE '%WD%') OR
     (s.sptypea LIKE '%CV%') OR
     (s.zbsubclass LIKE '%WD%') OR
     (s.zbsubclass LIKE '%CV%') OR
     (s.flag LIKE 'D%') OR                  -- likely WD
     (s.flag LIKE 'd%') OR                  -- likely SD
 -- TARGETing from SpecObj (primTarget should be same as targettype)
     ((t.primTarget & 0x000A0000) > 1) OR    -- WD or CATY_VAR
     ((t.secTarget & 0x00000200) >1) OR    -- HOT_STD
     ((t.seguePrimTarget & 0x000A0000) > 1) OR    -- WD or CATY_VAR
     ((t.segueSecTarget & 0x00000200) >1)    -- HOT_STD
  
ORDER by t.plate, t.mjd, t.fiberID

\end{lstlisting}

\end{document}